\documentclass[useAMS,usenatbib,usegraphicx]{mn2e}

\title[New sdB binaries]{Two new hot subdwarf binaries in the {\it GALEX} survey}
\author[A. Kawka, et al.]{A. Kawka$^{1,2}$\thanks{E-mail: kawka@sunstel.asu.cas.cz (AK); vennes@sunstel.asu. cas.cz (SV); pnemeth@fit.edu (PN); kraus@sunstel.asu.cas.cz (MK); kubat@sunstel.asu.cas.cz (JK)}, S. Vennes$^{1,2}$\footnotemark[1], P. N\'emeth$^3$\footnotemark[1], M. Kraus$^1$\footnotemark[1], J. Kub\'at$^1$\footnotemark[1]\\
$^{1}$Astronomick\'y \'ustav, Akademie v\v{e}d \v{C}esk\'e republiky, 
Fri\v{c}ova 298, CZ-251 65 Ond\v{r}ejov, Czech Republic\\
$^{2}$Visiting Astronomer, Kitt Peak National Observatory, National Optical Astronomy Observatory, which is operated by the\\
Association of Universities for Research in Astronomy (AURA) under cooperative agreement with the National Science Foundation.\\
$^{3}$Department of Physics and Space Sciences, Florida Institute of Technology,
150 West University Boulevard, Melbourne,\\ FL 32901-6975, USA}

\begin{document}

\date{Accepted;  Received ;} 

\pagerange{\pageref{firstpage}--\pageref{lastpage}} \pubyear{2010}

\maketitle

\label{firstpage}

\begin{abstract}
We report the discovery of two new hot, hydrogen-rich subdwarfs (sdB) in close 
binary systems. The hot subdwarfs, GALEX~J0321$+$4727 and GALEX~J2349$+$3844, were 
selected from a joint optical-ultraviolet catalogue of hot sub-luminous stars 
based on GSC2.3.2 and the {\it Galaxy Evolution Explorer} all-sky survey. Using 
high-dispersion spectra of the H$\alpha$ core obtained using the 2m telescope 
at Ond\v{r}ejov Observatory we measured the radial velocities 
of the sdB primaries and determined orbital periods of $0.26584\pm0.00004$ days 
and $0.46249\pm0.00007$ days for GALEX~J0321$+$4727 and GALEX~J2349$+$3844, 
respectively. The time series obtained from the Northern Sky Variability Survey 
with an effective wavelength near the R band show that GALEX~J0321$+$4727 is a 
variable star ($\Delta m=0.12$ mag) while no significant variations are 
observed in GALEX~J2349$+$3844. The period of variations in GALEX~J0321$+$4727 
coincides with the orbital period and the variability is probably caused by a 
reflection effect on a late-type secondary star. Lack of photometric variations 
in GALEX~J2349$+$3844 probably indicates that the companion is a white dwarf 
star. Using all available photometry and spectroscopy, we measured the 
atmospheric properties of the two sdB stars and placed limits on the mass and 
luminosity of the companion stars.
\end{abstract}


\section{Introduction}

Subdwarf B (sdB) stars are helium core burning stars with very thin 
hydrogen envelopes that lie at the blue end of the horizontal branch and hence
are identified with the extreme horizontal branch (EHB) stars \citep[see a recent review by][]{heb2009}.
\citet{dcr1996} showed that a high mass-loss rate on the red giant branch
produces a thin hydrogen envelope and prevents the star from ascending the
asymptotic giant branch. 
The evolution of single EHB stars from zero-age to helium exhaustion may be 
followed on a series of tracks narrowly centred on 0.475 $M_\odot$\citep{dor1993}.  
After helium exhaustion these objects evolve directly onto
the white dwarf sequence.

On the other hand, following the original proposal of \citet{men1976} for the
formation of sdB stars through binary evolution, it has been found that
a significant fraction of these stars reside
in close binary systems \citep[e.g.,][]{max2001,mor2003}.
The onset of a common envelope phase or Roche lobe overflow contributes to the
removal of the hydrogen envelope and directs the star toward the EHB.
\citet{han2002,han2003} propose three formation channels for the formation
of sdB stars through binary interaction, either involving common envelope (CE) phases,
episodes of Roche lobe overflow (RLOF), or the merger of two helium white dwarfs.
The CE scenario involving primary stars that experience a  
helium flash accompanied by a low-mass or white dwarf secondary star is expected
to create short-period binaries ($\log{P(d)}\approx -1$ to 1) and a
final primary mass distribution narrowly centred on 0.46$M_\odot$.
The CE scenario with primary stars massive enough to avoid a helium flash
is expected to achieve a much lower final mass for the primary (0.33-0.35 $M_\odot$).
On the other hand, the RLOF scenario creates longer period binaries and a wider distribution
of primary final masses. Studies of the binary components, and an
estimate of the frequency of such systems are required to constrain these models and
determine the relative contribution of these formation channels to the sdB population.

In this context, we have initiated a program to identify new hot
subdwarf candidates (Vennes et al., in preparation). 
We combined ultraviolet photometric measurements from the {\it Galaxy Ultraviolet Explorer} ({\it GALEX}) all-sky survey 
and photographic visual magnitudes from the Guide Star Catalog, Version 2.3.2 (GSC2.3.2), to build a list of blue stellar candidates, while
follow-up spectroscopic measurements further constrained the properties of the candidates. In Section 2,
we present available spectroscopic and photometric measurements of two new, hot sdB stars. 
Section 3 presents our analysis of the radial velocity measurements, and in
Section 4 we constrain the properties of the binary components. We summarise and conclude
in Section 5.

\section{Observations}

The ultraviolet sources 
GALEX~J234947.7$+$384440 (hereafter GALEX~J2349$+$3844)
and GALEX~J032139.8+472716 (hereafter GALEX~J0321$+$4727) were originally selected
based on the colour index $NUV-V<0.5$ and the brightness limit $NUV<14$, where $NUV$ is the {\it GALEX} near ultraviolet
bandpass, and $V$ is the GSC photographic magnitude.
The sources were also identified with optical counterparts in the Tycho-2
catalogue \citep{hog2000} and the Third U.S. Naval Observatory CCD Astrograph Catalog \citep[UCAC3,][]{zac2010}.
GALEX~J0321$+$4727 is also known as a disqualified member (No. 488) of the cluster Melotte 20 \citep[see][]{mer2008,van2009}.
\citet{hec1956} listed a spectral type of B7 
but excluded it as a possible cluster member based on its proper motion.
GALEX~J2349$+$3844 was independently identified in the
First Byurakan Survey of Blue Stellar Objects as FBS~2347+385 \citep{mic2008}.

\begin{table}
\centering
\begin{minipage}{\columnwidth}
\caption{Astrometry and photometry \label{tbl_phot}}
\begin{tabular}{@{}lcc@{}}
\hline
     & J0321$+$4727 & J2349$+$3844 \\
\hline
RA (2000)     & 03 21 39.629     & 23 49 47.645     \\
Dec (2000)    & +47 27 18.79     & +38 44 41.57     \\
$\mu_{\alpha}\cos{\delta}$ (Tycho-2) & 57.2$\pm$1.9 mas yr$^{-1}$ & $-$7.5$\pm$3.5 mas yr$^{-1}$ \\
$\mu_{\delta}$ (Tycho-2) & $-$8.5$\pm$1.8 mas yr$^{-1}$ & 1.6$\pm$3.2 mas yr$^{-1}$ \\
$\mu_{\alpha}\cos{\delta}$ (UCAC3) & 58.0$\pm$1.0 mas yr$^{-1}$ & $-$4.0$\pm$2.3 mas yr$^{-1}$ \\
$\mu_{\delta}$ (UCAC3) & $-$8.4$\pm$1.0 mas yr$^{-1}$ & $-$1.4$\pm$1.3 mas yr$^{-1}$ \\
{\it FUV}     & $12.441\pm0.019$ & $11.261\pm0.021$ \\
{\it NUV}     & $11.913\pm0.007$ & $11.310\pm0.005$ \\
$B$           & $11.53\pm0.11$   & $11.66\pm0.11$   \\
$V$           & $11.72\pm0.16$   & $11.72\pm0.15$   \\
2MASS {\it J} & $11.807\pm0.023$ & $12.040\pm0.024$ \\
2MASS {\it H} & $11.859\pm0.030$ & $12.156\pm0.031$ \\
2MASS {\it K} & $11.893\pm0.029$ & $12.184\pm0.024$ \\
\hline
\end{tabular}
\end{minipage}
\end{table}

\subsection{Photometry and astrometry}

We extracted the ultraviolet photometry from the 
{\it GALEX} all-sky survey using CasJobs at the Multimission Archive at STScI (MAST). {\it GALEX} obtained photometric 
measurements in the FUV and NUV bands with effective wavelengths of 1528 \AA\ and 2271 \AA, 
respectively. We corrected the photometry for non-linearity 
\citep{mor2007}. Although the statistical errors are
small ($<0.02$ mag) we estimate the total errors to be $\ga 0.2$ mag because of large uncertainties in the
linearity corrections for bright sources \citep[see][]{mor2007}. In the present case, the $NUV$ measurements are more
reliable than the $FUV$ measurements.
We also obtained infrared photometry from the Two Micron All Sky 
Survey \citep[{\it 2MASS},][]{skr2006} and optical photometry from the Tycho catalogue\footnote{Accessed at VizieR \citep{och2000}.}. 
We transformed the Tycho $B_T$ and $V_T$ photometric magnitudes
to the Johnson $B$ and $V$ magnitudes using the recommended transformation equations
\citep{per1997}. 

Table~\ref{tbl_phot} lists the available photometry 
for the two sources and astrometric measurements from Tycho-2 and UCAC3. The tabulated coordinates (epoch and equinox 2000)
are the averages of the Tycho-2 and UCAC3 coordinates. Both optical counterparts lie within $\sim 3\arcsec$ of the
ultraviolet sources. 


Finally, we have extracted photometry from the Northern Sky Variability Survey 
(NSVS). The photometric bandpass is very broad, ranging from 4500 to 10000 \AA, 
with an effective wavelength close to the Johnson $R$ band \citep{woz2004}. 
The modified julian dates supplied by NRVS were converted to the barycentric 
julian dates. The time series comprise 173 and 240 good measurements for 
GALEX~J0321$+$4727 and GALEX~J2349$+$3844, respectively, and allow the 
examination of our objects for variability. We also obtained a photometric 
series (167 good measurements) of the known eclipsing and variable sdB+dM 
binary 2M~1533$+$3759 \citep{for2010} to test our methodology. 

\subsection{Spectroscopy}

We observed GALEX~J2349$+$3844 and 
GALEX~J0321$+$4727 using the spectrograph at the coud\'e focus of the 2m 
telescope at Ond\v{r}ejov Observatory \citep{sle2002}. We obtained the spectroscopic series using the 830.77
lines per mm grating with a SITe $2030\times 800$ CCD that delivered
a spectral resolution $R = 13\, 000$ and a spectral range from 6254 \AA\ to 6763 \AA. 
The exposure time for both targets is 30 minutes, with each exposure immediately 
followed by a ThAr comparison arc.
The fast rotating B star HR 7880 was observed each 
night to help remove telluric features from the spectra. 
We verified the stability of the wavelength scale by measuring the wavelength centroids
of O{\sc i} sky lines. The velocity scale remained stable within 1\,km~s$^{-1}$.

We also obtained two low dispersion spectra of GALEX~J0321$+$4727 using the
R-C spectrograph attached to the 4m telescope at Kitt Peak National Observatory
(KPNO) on UT 2010 March 23. 
We used the KPC-10A grating (316 lines/mm) with the 
WG360 order blocking filter. The slitwidth was set to 1.5 arcseconds to provide
a resolution of FWHM $= 5.5$\AA. A HeNeAr comparison spectrum was obtained following
the target spectrum.
We exposed GALEX~J0321$+$4727 for 60 and 180 s and we co-added the spectra
weighted by the exposure times. 
All spectra were reduced using standard procedures within IRAF.

\section{Binary Parameters}

\subsection{Radial velocity variations}

We measured the radial velocities by fitting a Gaussian function to the 
H$\alpha$ core and by measuring the shifts relative to the rest wavelength. 
The shifts were then converted into radial velocities and adjusted to the 
solar system barycentre. Tables~\ref{tbl_gal0321} and \ref{tbl_gal2349} lists 
the barycentric julian dates, radial velocities and the spectra signal-to-noise 
ratios for GALEX~J0321$+$4727 and GALEX~J2349$+$3844,
respectively. The accuracy of individual measurements varied from 1
\,km~s$^{-1}$ in high signal-to-noise spectra to 10\,km~s$^{-1}$
in lower quality spectra.

\begin{table}
\centering
\begin{minipage}{\columnwidth}
\caption{Radial velocities of GALEX~J0321$+$4727. \label{tbl_gal0321}}
\begin{tabular}{@{}ccc|ccc@{}}
\hline
BJD      & v              & S/N & BJD & v & S/N \\
(2455000+) & (km s$^{-1}$) &  & (2455000+) & (km s$^{-1}$) & \\
\hline
45.54317 &  +24.9 & 29 & 75.61381 &  +38.6 & 13 \\
45.57495 &  +61.4 & 32 & 76.40849 &  +66.1 & 20 \\
59.49449 & +119.6 & 25 & 76.43166 &  +86.7 & 21 \\
62.58712 &  +58.9 & 15 & 76.45473 & +112.2 & 22 \\
62.61018 &  +95.5 & 10 & 76.47782 & +120.4 & 22 \\
63.55702 &  +45.3 & 10 & 76.49589 & +120.9 & 11 \\
63.57998 &  +13.8 & 15 & 76.52687 &  +95.7 & 14 \\
63.60293 &  +15.1 & 17 & 76.54989 &  +80.6 & 15 \\
63.61545 &  +13.7 & 12 & 76.57303 &  +49.1 & 14 \\
75.41035 & +143.4 & 15 & 84.56626 &  +35.5 & 13 \\
75.43322 & +131.9 & 18 & 84.59055 &   +9.4 & 26 \\
75.48607 &  +82.6 & 15 & 84.61363 &  +12.2 & 24 \\
75.50993 &  +46.0 & 13 & 84.63684 &  +35.0 & 25 \\
75.53328 &  +14.5 & 16 & 98.38356 &  +18.2 &  8 \\
75.59048 &  +28.1 & 13 &          &        &    \\
\hline
\end{tabular}
\end{minipage}
\end{table}

\begin{table}
\centering
\begin{minipage}{\columnwidth}
\caption{Radial velocities of GALEX~J2349$+$3844. \label{tbl_gal2349}}
\begin{tabular}{@{}ccc|ccc@{}}
\hline
BJD        & v             & S/N & BJD    & v & S/N \\
(2455000+) & (km s$^{-1}$) & & (2455000+) & (km s$^{-1}$) & \\
\hline
45.48574 & -32.6 & 21 & 75.56357 &  -4.7 & 12 \\
45.50700 &  +2.0 & 22 & 76.31684 & -47.6 & 17 \\
59.40594 & +29.0 & 25 & 76.33994 & -71.0 & 18 \\
62.49909 & -62.8 &  8 & 76.36123 & -82.4 & 16 \\
63.46112 & -76.3 & 15 & 76.38463 & -79.2 & 20 \\
63.48409 & -70.0 & 18 & 84.28986 & -74.6 & 25 \\
63.51335 & -41.2 & 17 & 84.31270 & -61.3 & 26 \\
63.53634 & -10.2 & 17 & 84.33575 & -38.1 & 27 \\
68.42225 & -19.7 & 14 & 84.35898 &  -8.4 & 25 \\
68.46553 & -55.4 & 12 & 84.38179 & +38.6 & 27 \\
69.35158 & -31.8 & 14 & 84.40464 & +54.1 & 27 \\
69.37813 & -52.8 & 19 & 84.42747 & +78.3 & 30 \\
69.40226 & -67.9 &  9 & 84.45030 & +86.0 & 23 \\
70.50880 & +23.4 & 18 & 84.47349 & +86.9 & 19 \\
70.53208 & +56.3 & 27 & 84.49679 & +86.4 & 22 \\
74.53854 & -97.6 & 13 & 84.52089 & +68.0 & 24 \\
75.31818 & +26.7 & 18 & 84.54375 & +53.4 & 25 \\
75.34117 &  +2.9 & 15 & 98.29111 & +59.4 & 17 \\
75.36393 & -20.4 & 18 & 98.31403 & +76.7 & 16 \\
75.38821 & -38.7 & 15 & 98.33709 & +90.4 & 17 \\
75.46518 & -81.3 & 11 & 98.36017 & +78.5 & 10 \\
\hline
\end{tabular}
\end{minipage}
\end{table}

\begin{figure}
\includegraphics[width=\columnwidth]{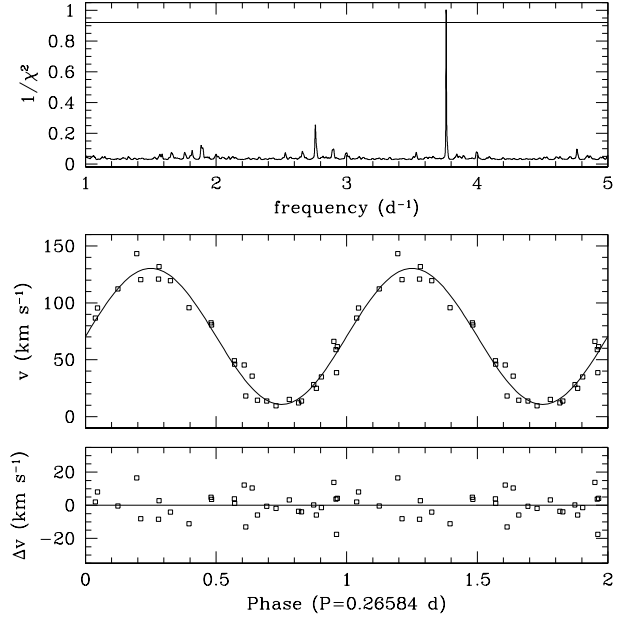}
\caption{({\it Top}) Period analysis of GALEX~J0321$+$4727 showing a single 
significant period. ({\it Middle}) Radial velocity measurements folded on the 
orbital period and best-fit sine curve. ({\it Bottom}) Residual of the 
velocities relative to the best-fit sine curve.\label{fig_vel_gal0321}}
\end{figure}

\begin{figure}
\includegraphics[width=\columnwidth]{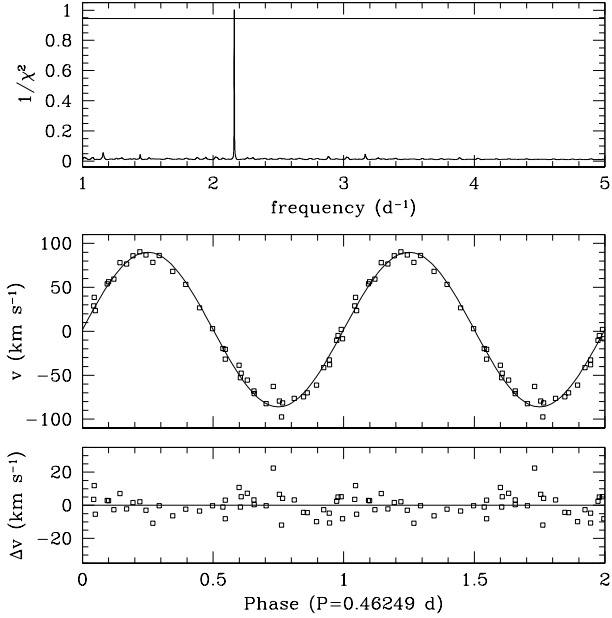}
\caption{Same as Figure~\ref{fig_vel_gal0321} but for GALEX~J2349$+$3844. 
\label{fig_vel_gal2349}}
\end{figure}

The orbital parameters were determined by 
fitting to the velocity series a sinusoidal function of the form 
\begin{displaymath}
v(t) = \gamma + K\sin{(2\pi[t-T_0]/P)},
\end{displaymath}
where $P$ is the period, $\gamma$ is the systemic velocity, $K$ is the
velocity semi-amplitude, and $T_0$ is the initial epoch. The initial epoch 
$T_0$ corresponds to the inferior conjunction of the sdB ($\Phi = 0$).
We applied a $\chi^2$ minimisation technique with each velocity measurement 
weighted proportionally to the signal-to-noise ratio achieved in the 
corresponding spectrum. Figures~\ref{fig_vel_gal0321} and \ref{fig_vel_gal2349} 
show the periodograms and best-fit radial velocity curves for 
GALEX~J0321$+$4727 and GALEX~J2349$+$3844, respectively. The residual to the 
best-fit radial curve is $\approx 6$ km\,s$^{-1}$ for both data sets.
Table~\ref{tbl_bin} lists the corresponding binary parameters and the 
calculated mass functions. The measured radial velocities were also employed to 
apply doppler corrections to individual spectra and build phase-averaged 
spectra for each star (Section 4).

\begin{table}
\centering
\begin{minipage}{\columnwidth}
\caption{Binary parameters. \label{tbl_bin}}
\begin{tabular}{@{}lcc@{}}
\hline
Parameter & J0321$+$4727 & J2349$+$3844 \\
\hline
Period (d)                   & $0.26584\pm0.00004$ & $0.46249\pm0.00007$ \\ 
$T_0$ (BJD 2455000+)         & $45.582\pm0.011$    & $45.511\pm0.013$         \\
K (km~s$^{-1}$)              & $59.8\pm4.5$        & $87.9\pm2.2$        \\
$\gamma$ (km~s$^{-1}$)       & $70.5\pm2.2$        & $2.0\pm1.0$         \\
$f(M_{\rm sec})$ ($M_\odot$) & $0.00589\pm0.00015$ & $0.03254\pm0.00044$ \\
\hline
\end{tabular}
\end{minipage}
\end{table}

Our new systemic velocity for GALEX~J0321$+$4727 ($\gamma=70.5$ km~s$^{-1}$) also clearly rules out membership
to the cluster Melotte 20 \citep[$\gamma=-1.4$ km~s$^{-1}$,][]{mer2008}.

\subsection{Photometric variations}

\begin{figure}
\includegraphics[width=\columnwidth]{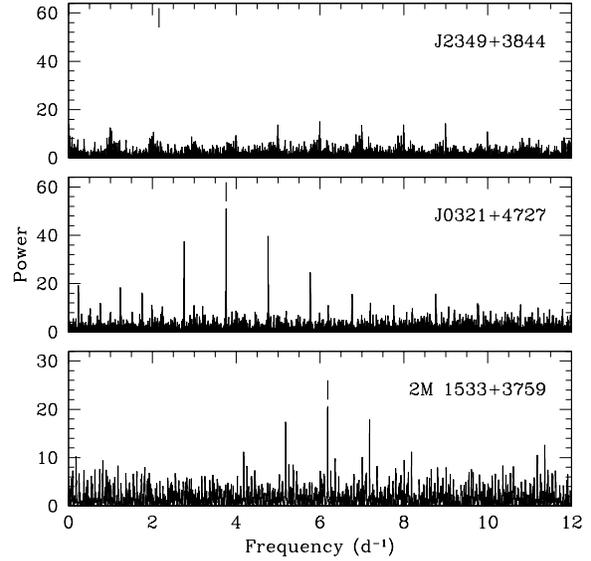}
\caption{Lomb periodograms of GALEX~J0321$+$4727 and GALEX~J2349$+$3844,
and the test target 2M~1533$+$3759.}
\label{fig_fft}
\end{figure}

\begin{figure}
\includegraphics[width=\columnwidth]{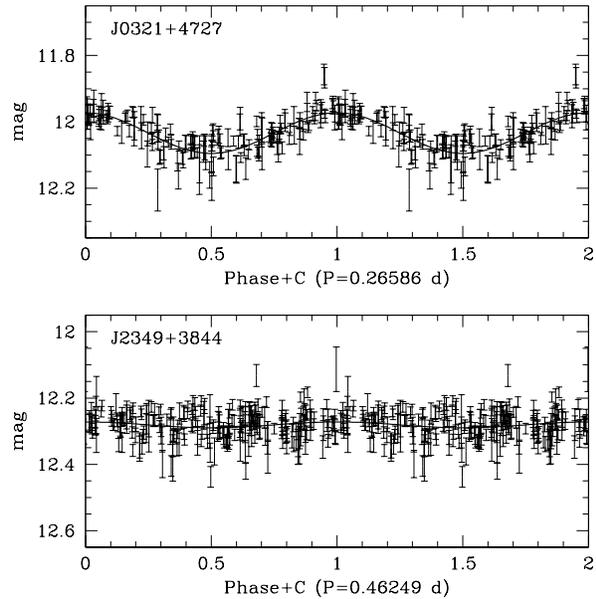}
\caption{NSVS lightcurves of GALEX~J0321$+$4727 and GALEX~J2349$+$3844 compared to best-fit 
sine curves with semi-amplitudes of $0.061\pm0.004$ and $0.009\pm0.004$ mag, respectively.}
\label{fig_phot}
\end{figure}

We investigated possible variability in GALEX~J0321$+$4727 and GALEX~J2349$+$3844
using the NSVS light curves. We analyse the light curves using a Lomb periodogram
for unevenly sampled time series \citep{pre1992}. The power spectra (Fig.~\ref{fig_fft})
show a peak signal at a frequency close to the orbital period in GALEX~J0321$+$4727 (Table~\ref{tbl_bin})
and 2M~1533$+$3759 \citep{for2010}, but not in GALEX~J2349$+$3844 (Table~\ref{tbl_bin}). We estimated the probability of
a given frequency relative to the probability of the peak frequency following \citet{pre1992} and determined
the 1$\sigma$ (66\%) error bars on the period of the photometric variations for GALEX~J0321$+$4727:
\begin{displaymath}
P=0.26586\pm0.00003\ {\rm d},
\end{displaymath}
and 2M~1533$+$3759:
\begin{displaymath}
P=0.16177\pm0.00003\ {\rm d}.
\end{displaymath}
For both stars, the period of photometric variations is equal, within error bars, to the measured orbital period,
thereby validating the method.
Figure~\ref{fig_phot} shows the NSVS light curves for GALEX~J0321$+$4727 and GALEX~J2349$+$3844 folded
on the orbital period with an arbitrary phase adjustment so that $\Phi=0$ corresponds to the
inferior conjunction of the primary star (sdB). The distant NSVS epoch (~1999) precluded phasing with the current
orbital ephemeris (Table~\ref{tbl_bin}). The light curve of GALEX~J0321$+$4727 is fitted with the sine curve:
\begin{displaymath}
m = (12.034\pm0.003) + (0.061\pm0.004)\ \sin{2\pi\Phi},
\end{displaymath}
that we interpret as a reflection effect on a late-type secondary star (Section 4). The variations in GALEX~J2349$+$3844
are not significant with a mean magnitude of $<m>=12.281\pm0.003$ and a semi-amplitude of $\Delta m/2=0.009\pm0.004$.

\section{Properties of the components}

Our analysis of the spectroscopic observations of the primary stars is based on
a grid of non-LTE models and synthetic spectra calculated using TLUSTY/SYNSPEC 
\citep{hub1995,lan1995}. The grid covers the effective temperature from 
$T_{\rm eff}=21000$ to 35000 K (in steps of 1000 K), the surface gravity from 
$\log{g}=4.75$ to 6.25 (in steps of 0.25), and the helium abundance from 
$\log{(n_{\rm He}/n_{\rm H})} = -4.0$ to $-1.0$ (in steps of 0.5). The neutral 
hydrogen and helium atoms include 9 ($n\le9$) and 24 ($n\le8$) energy levels,
respectively, and the ionized helium atom includes 20 ($n\le20$) energy levels.

\subsection{Spectral energy distribution}

\begin{figure}
\includegraphics[width=\columnwidth]{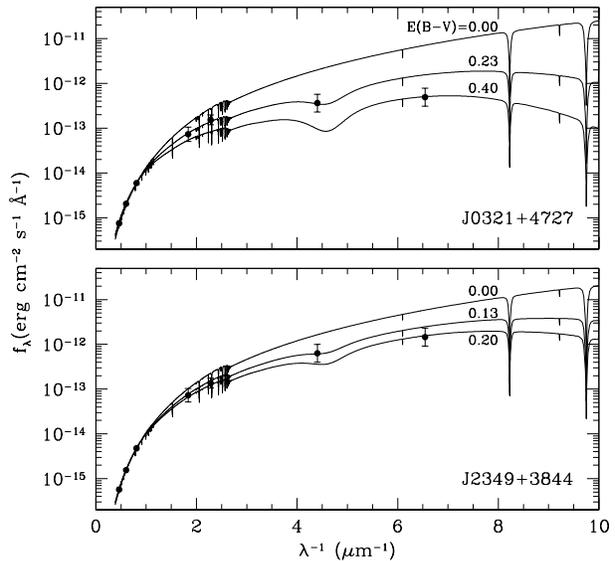}
\caption{({\it Top}) Spectral energy distribution of GALEX~J0321$+$4727 
compared to a model spectrum that was corrected for interstellar extinction 
assuming $E(B-V) = 0.0, 0.23, 0.4$, and $R_V = 3.2$. ({\it Bottom}) Same but 
for GALEX~J2349$+$3844 and assuming $E(B-V) = 0.0, 0.13, 0.2$.
\label{fig_sed_gal2349}}
\end{figure}

Figure~\ref{fig_sed_gal2349} shows a preliminary analysis of the atmospheric 
properties of the two sdB stars using their observed spectral energy 
distribution (SED) from the infrared to the ultraviolet, and using 
representative sdB models at $T_{\rm eff}=29000$ K, $\log{g}=5.5$, and 
$\log{(n_{\rm He}/n_{\rm H})}=-2.65$ (GALEX~J0321$+$4727) and $-3.25$ 
(GALEX~J2349$+$3844). The model corresponds to a hot sdB star with absolute 
magnitude $M_V=4.1$ and a canonical mass of $0.47\,M_\odot$ 
\citep[see][]{dor1993}. We corrected the model spectra for interstellar 
extinction using a variable extinction coefficient $E(B-V)$ and a parametrised 
extinction law \citep[$R=3.2$,][]{car1989}. GALEX~J0321$+$4727 lies close to 
the plane of the Galaxy ($l=147.5, b=-8.1$) and the total extinction in the
line of sight is $E(B-V) = 0.61$ \citep{sch1998}. Excluding the FUV band, the 
observed SED is well matched by the model assuming $E(B-V) = 0.23$. A much 
higher coefficient ($\approx 0.4$) is required to match the FUV band although 
the accuracy of the FUV photometry is most probably affected by non-linearity 
(see Section 2.1). GALEX~J2349$+$3844 is located lower below the plane 
($l=110.0, b=-22.6$) and the total extinction is lower $E(B-V) = 0.17$. In this 
case, the SED is well matched assuming $E(B-V) = 0.13$. These estimates appear 
reasonable if we locate both stars at a distance of $\sim 330$ pc by assuming 
$M_V\sim4.1$ and $V=11.7$ for both stars. Taking the scale height for dust in
the Galactic plane as $h\approx 150$ pc, the total distance accross the dust 
layer is $\approx 1060$ and 390 pc at $|b|=8.1$ and 22.6$^\circ$, respectively, 
so that the path toward GALEX~J0321$+$4727 covers $\sim31$\% of the total 
distance and $\sim85$\% for GALEX~J2349$+$3844. According to this simple 
calculation, the scaled $E(B-V)$ indices are predicted to be $0.19$ and $0.15$ 
for GALEX~J0321$+$4727 and GALEX~J2349$+$3844, 
respectively, and are similar to our estimates based on the SED.

Taking into account the effect of interstellar reddening, the entire spectral energy distribution
of both systems
is dominated by the sdB stars.

\subsection{Line profile analysis}

We measured the sdB effective temperature ($T_{\rm eff}$), surface gravity
($\log{g}$), and helium abundance ($\log{(n_{\rm He}/n_{\rm H})}$) by
fitting the observed line profiles with our grid of model spectra.
We employed $\chi^2$ minimisation techniques to find the best-fit parameters
and draw contours at 66, 90, and 99\% significance. 
The model spectra were convolved with Gaussian profiles 
with a $FWHM = 5.5$ \AA\ 
for the analysis of the KPNO spectra, while we adopted a $FWHM= 0.66$ \AA\ that includes
the effect of orbital smearing for the Ond\v{r}ejov coud{\'e} spectra.

\begin{figure}
\includegraphics[width=\columnwidth]{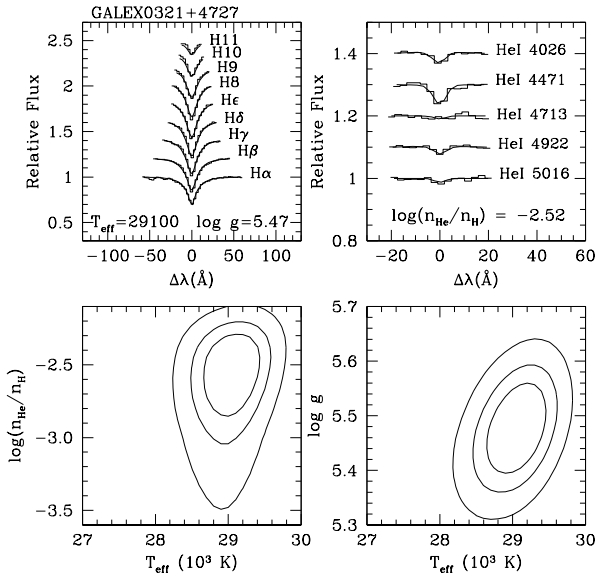}
\caption{Model atmosphere analysis of the low-dispersion KPNO spectrum of 
GALEX~J0321$+$4727. The {\it top} panels show the line profiles and best fit 
models, and the {\it lower} panels show the $\chi^2$ contours drawn at 66, 90, 
and 99\%.}
\label{fig_spec}
\end{figure}

First we analysed the KPNO spectrum of GALEX~J0321$+$4727 (Fig.~\ref{fig_spec}).
We included in the analysis the Balmer line spectrum from H$\alpha$ to H11 and 
five blue He{\sc i} lines normally dominant in sdB stars. We repeated the
analysis with H$\alpha$ excluded and obtained the same atmospheric parameters 
as the analysis that included H$\alpha$.
The mid-exposure time of the co-added KPNO spectrum is BJD~2455278.59501 
corresponding to an orbital phase $\Phi=0.51\pm0.03$ or close to the inferior 
conjunction of the secondary star. This phase also corresponds to minimum 
contamination to the sdB spectrum due to the reflection effect. 

\begin{figure}
\includegraphics[width=\columnwidth]{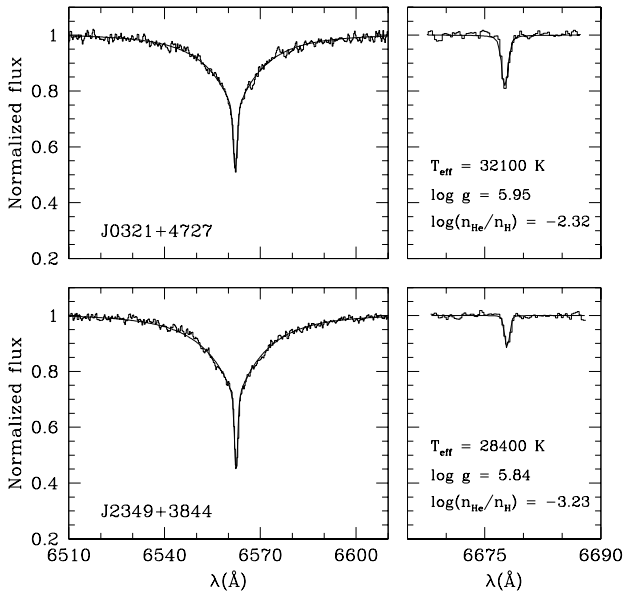}
\caption{Co-added coud{\'e} spectra of GALEX~J0321$+$4727 ({\it top}) and 
GALEX~J2349$+$3844 ({\it bottom}) showing the H$\alpha$ ({\it left}) and HeI 
$\lambda 6678$ \AA\ ({\it right}) lines and the best-fit models. 
\label{fig_fit}}
\end{figure}

Next, we analysed phase-resolved H$\alpha$ spectra of GALEX~J0321$+$4727 to
take into account the light contamination due to the reflection effect on the 
temperature measurements. This effect was notable in the analysis of the 
similar systems such as HS2333+3927 \citep{heb2004} and 2M~1533+3759 
\cite{for2010}. Variations of $\approx 6000$ K were observed in HS2333+3927, 
while weaker variations of $\approx 1000$ K were observed in 2M~1533+3759. To 
investigate this effect in GALEX~J0321$+$4727 we built three spectra inclusive 
of phases $0.0-1.0$ (average), $0.35-0.65$ (minimum reflection effect), and
$0.85-0.15$ (maximum reflection effect). 
Figure~\ref{fig_fit} shows our analysis of the 
H$\alpha$ and He I$\lambda 6678$ \AA\ line profiles in the co-added ($0.0-1.0$) coud{\'e} 
spectra of GALEX~J0321$+$4727 and GALEX~J2349$+$3844.

Table~\ref{tbl_sdb} summarises our measurements of the hot subdwarf atmospheric 
parameters. Our measurements show that variability in GALEX~J0321$+$4727 is 
affecting the temperature and abundance measurements but that the surface 
gravity does not vary significantly. In our limited phase resolution, the 
peak-to-peak temperature variation reaches $\approx 4000$ K. The Ond\v{r}ejov 
coud{\'e} and KPNO temperature measurements at phase 0.5 agree but the surface 
gravity measurements differ by 0.5 dex. This is most likely due to a systematic 
effect in the model spectra themselves with the H$\alpha$ line profile analysis 
overestimating the surface gravity relative to an analysis involving the 
complete series. The shape and strength of the upper Balmer 
lines are very sensitive to surface gravity and offer a more reliable surface 
gravity diagnostics. In summary we estimate the
parameters representative of the sdB GALEX~J0321$+$4727 at phase 0.5. The
effective temperature and helium abundance were estimated by taking
the weighted average of the results from the H$\alpha$-H11 and H$\alpha$ 
(0.35 - 0.65) fits and for the surface gravity we adopted the result from the
H$\alpha$-H11 analysis:
\begin{displaymath}
T_{\rm eff}=29200\pm300,\ \log{g}=5.5\pm0.1,\ 
\end{displaymath}
and
\begin{displaymath}
\log{n_{\rm He}/n_{\rm H}}=-2.6\pm0.1.
\end{displaymath}
Applying a correction of $-0.5\pm0.2$ to the surface gravity measurement of GALEX~J2349$+$3844
that is based on H$\alpha$ alone we conservatively estimate the sdB parameters:
\begin{displaymath}
T_{\rm eff}=28400\pm400,\ \log{g}=5.4\pm0.3,\ 
\end{displaymath}
and
\begin{displaymath}
\log{n_{\rm He}/n_{\rm H}}=-3.2\pm0.1.
\end{displaymath}

Our analysis assumes a hydrogen/helium composition and the inclusion of heavy
elements in the model atmospheres is likely to affect our results 
\citep[see][]{heb2000,ede2003}. A self consistent analysis including abundance 
measurements \citep[e.g.,][]{oto2006,ahm2007} awaits high signal-to-noise and
resolution ultraviolet and optical spectroscopy.

\begin{table}
\centering
\begin{minipage}{\columnwidth}
\caption{Measurements. \label{tbl_sdb}}
\begin{tabular}{ccccc}
\hline
Range & Phase & $T_{\rm eff}$ & $\log{g}$ & $\log{n_{\rm He}/n_{\rm H}}$ \\ 
      &       &   (K)         &  (cgs)    &                               \\
\hline
\multicolumn{5}{c}{J0321$+$4727} \\
\hline
H$\alpha$ & 0.85-0.15  & 33750$\pm$350 & 5.88$\pm$0.07 & $-$2.10$\pm$0.10\\
H$\alpha$ & 0.00-1.00  & 32100$\pm$250 & 5.95$\pm$0.05 & $-$2.32$\pm$0.07\\
H$\alpha$ & 0.35-0.65  & 29550$\pm$650 & 5.98$\pm$0.09 & $-$2.68$\pm$0.20\\
H$\alpha$-H11 & 0.50   & 29100$\pm$350 & 5.47$\pm$0.08 & $-2.52_{-0.31}^{+0.22}$\\ 
\hline
\multicolumn{5}{c}{J2349$+$3844} \\
\hline
H$\alpha$ & 0.00-1.00  & 28400$\pm$400 & 5.84$\pm$0.06 & $-$3.23$\pm$0.09 \\
\hline
\end{tabular}
\end{minipage}
\end{table}

\subsection{Nature of the companions}

Using the mass function (Table~\ref{tbl_bin})
and assuming a mass of $0.5\ M_\odot$ for the hot subdwarf GALEX~J0321$+$472,
we calculate a minimum secondary mass of $0.13\ M_\odot$ that 
corresponds to a spectral type of M5 \citep{kir1994}. Assuming an absolute $J$
magnitude of 3.9 for the hot subdwarf, a M5 star with 
$M_J = 8.8$ would be outshone by the hot subdwarf. 

The NSVS lightcurve
shows GALEX~J0321$+$4727 to be variable and
the search for a period in the photometric data 
resulted in a best-period corresponding to the orbital period. The
observed variations are most probably caused by irradiation of the atmosphere of the cool
companion by the hot subdwarf. 
Using the observed semi-amplitude of
0.061 mag we may constrain the binary parameters further. We 
estimated the system inclination and secondary mass using the reflection model of \citet{max2002} and
assuming two different masses for
the sdB star. For
a sdB mass of $0.4\ M_\odot$, the inclination is predicted to be between
63$^\circ$ and 71$^\circ$ and the secondary mass between $0.124\ M_\odot$ and 
$0.133\ M_\odot$. For a sdB mass of $0.5\ M_\odot$, we obtain an 
inclination ranging from 65$^\circ$ to 70$^\circ$, and a secondary mass between 
$0.143\ M_\odot$ and $0.149\ M_\odot$.

Again, using the mass function and assuming a mass of $0.5\ M_\odot$ for the 
hot subdwarf GALEX~J2349$+$3844, the minimum secondary mass is $0.27\ M_\odot$
that corresponds to a spectral type of M4 \citep{kir1994}. Assuming an absolute
$J$ magnitude of 5.6 for the hot subdwarf, a M4 spectral type star with 
$M_J = 8.6$ would also be outshone by the hot subdwarf.

However, the NSVS time series do not show variations down to a limit of
$\Delta m  = 0.009$. Illumination of a $0.3\ M_\odot$ star, which is the 
suggested mass at a high inclination, would cause a variation of 
$\Delta m \sim 0.4$ magnitudes. Lower inclinations would require larger
companions causing even larger variations that are incompatible with the 
observations. The lack of variability suggests that the companion is most 
likely a white dwarf \citep[see][]{max2004}.

\section{Summary and Conclusions}

We show that GALEX~J0321$+$4727 and GALEX~J2349$+$3844 are hot hydrogen-rich 
subdwarfs in close binaries. Based on a preliminary analysis of periodic light 
variations in GALEX~J0321$+$4727 we infer that its companion is a low-mass star 
($M\sim0.13\,M_\odot$). The secondary star in GALEX~J2349$+$3844 is probably a 
white dwarf with $M\ga 0.3\, M_\odot$. The two new systems are post-CE systems 
with a hot subdwarf primary. Their orbital periods locate them close to the 
peak of the period distribution for such systems \citep[see][]{heb2009}. A 
future study of GALEX~J0321$+$4727 will involve phase-resolved high 
signal-to-noise ratio spectroscopic and photometric observations aimed at 
resolving the nature of the companion. Searches for close binaries in the sdB 
population have a relatively high yield \citep[69\%, see][]{max2001},
and, therefore, we expect that many new systems remain to be discovered in our 
GALEX/GSC catalogue of EHB stars.

\section{acknowledgements}

S.V. and A.K. are supported by GA AV grant numbers IAA300030908 and IAA301630901, respectively, and by GA \v{C}R grant number P209/10/0967.
A.K. also acknowledges support from the Centre for Theoretical Astrophysics (LC06014).
Some of the data presented in this paper were obtained from the Multimission Archive at the Space Telescope Science Institute (MAST). STScI is operated by the Association of Universities for Research in Astronomy, Inc., under NASA contract NAS5-26555. Support for MAST for non-HST data is provided by the NASA Office of Space Science via grant NNX09AF08G and by other grants and contracts.

\label{lastpage}

\end{document}